\documentclass[10pt,preprint]{aastex}

\slugcomment{{\sc Accepted to ApJ:} April 2, 2012}

\shorttitle{Relevance of polyynyl-PAHs to astrophysics}

\shortauthors{Rouill\'e et al.}

\begin{document}

\title{On the relevance of polyynyl-substituted \\
    PAHs to astrophysics}

\author{G. Rouill\'e, M. Steglich, Y. Carpentier, C. J\"ager, F. Huisken}
\affil{Laboratory Astrophysics Group of the Max Planck Institute for Astronomy at the Friedrich Schiller University Jena, Institute of Solid State Physics, Helmholtzweg 3, D-07743 Jena, Germany}
\email{friedrich.huisken@uni-jena.de}

\author{Th. Henning}
\affil{Max Planck Institute for Astronomy, K\"onigstuhl 17, D-69117 Heidelberg, Germany}

\author{R. Czerwonka, G. Theumer, C. B\"orger, I. Bauer, and H.-J. Kn\"olker}
\affil{Department Chemie, Technische Universit\"at Dresden, Bergstrasse 66, D-01069 Dresden, Germany}

\begin{abstract}
We report on the absorption spectra of the polycyclic aromatic hydrocarbon (PAH) molecules anthracene, phenanthrene, and pyrene carrying either an ethynyl ($-$C$_2$H) or a butadiynyl ($-$C$_4$H) group. Measurements were carried out in the mid infrared at room temperature on grains embedded in CsI pellets and in the near ultraviolet at cryogenic temperature on molecules isolated in Ne matrices. The infrared measurements show that interstellar populations of polyynyl-substituted PAHs would give rise to collective features in the same way non-substituted PAHs give rise to the aromatic infrared bands. The main features characteristic of the substituted molecules correspond to the acetylenic CH stretching mode near 3.05~$\mu$m and to the almost isoenergetic acetylenic CCH in- and out-of-plane bending modes near 15.9~$\mu$m. Sub-populations defined by the length of the polyynyl side group cause collective features which correspond to the various acetylenic CC stretching modes. The ultraviolet spectra reveal that the addition of an ethynyl group to a non-substituted PAH molecule results in all its electronic transitions being redshifted. Due to fast internal energy conversion, the bands at shorter wavelengths are significantly broadened. Those at longer wavelengths are only barely affected in this respect. As a consequence, their relative peak absorption increases. The substitution with the longer butadiynyl chain causes the same effects with a larger magnitude, resulting in the spectra to show a prominent if not dominating $\pi$-$\pi^*$ transition at long wavelength. After discussing the relevance of polyynyl-substituted PAHs to astrophysics, we conclude that this class of highly conjugated, unsaturated molecules are valid candidates for the carriers of the diffuse interstellar bands.
\end{abstract}


\section{Introduction}

Infrared (IR) emission bands reveal the presence of polycyclic aromatic hydrocarbon (PAH) molecules in the interstellar medium (ISM) \citep{Leger84,Allamandola85}. These so-called aromatic infrared bands (AIBs) represent a collective fingerprint of the interstellar PAHs, which cannot be resolved to lead to the identification of individual molecules. As the relative intensities and profiles of the AIBs vary depending on the region where the bands are observed, different populations of PAHs are proposed to account for the variations. The populations are defined in terms of charge state, size distribution, presence of heterocycles and side groups, and clusterization state \citep[for a review, see][]{Tielens08}.

\citet{Duley09} measured IR spectra of synthetic carbon nanoparticles containing $sp$-bonded chains, both polyynes and cumulenes. Having found similarities between the laboratory spectra and interstellar IR emission bands, they conjectured the presence of PAH molecules carrying $sp$-bonded chains in IR emission regions. To date, this conjecture awaits verification. The failure to find the molecules in IR sources, however, would not demonstrate their absence from regions in which different conditions prevail and, therefore, another chemistry takes place, such as the diffuse ISM.

The search for molecules in the diffuse ISM is conducted by looking for absorption bands corresponding to their electronic transitions, which typically occur at near ultraviolet (UV) and visible (vis) wavelengths. Since little is known about the electronic spectra of PAHs carrying $sp$-bonded chains, it appears worthwhile to undertake a study of such species.

We report on the UV/vis absorption spectra of polyynyl-substituted PAHs isolated in Ne matrices at 6~K and on the IR spectra of these molecules obtained with samples embedded in salt pellets. Quantum chemical calculations have been performed to assist in the interpretation of the laboratory spectra. In the light of the new data, we discuss the relevance of polyynyl-substituted PAHs to astrophysics.

\section{Experimental and Computational Details}

\subsection{Molecules} \label{ex:species}

\begin{figure}
\epsscale{0.50}
\plotone{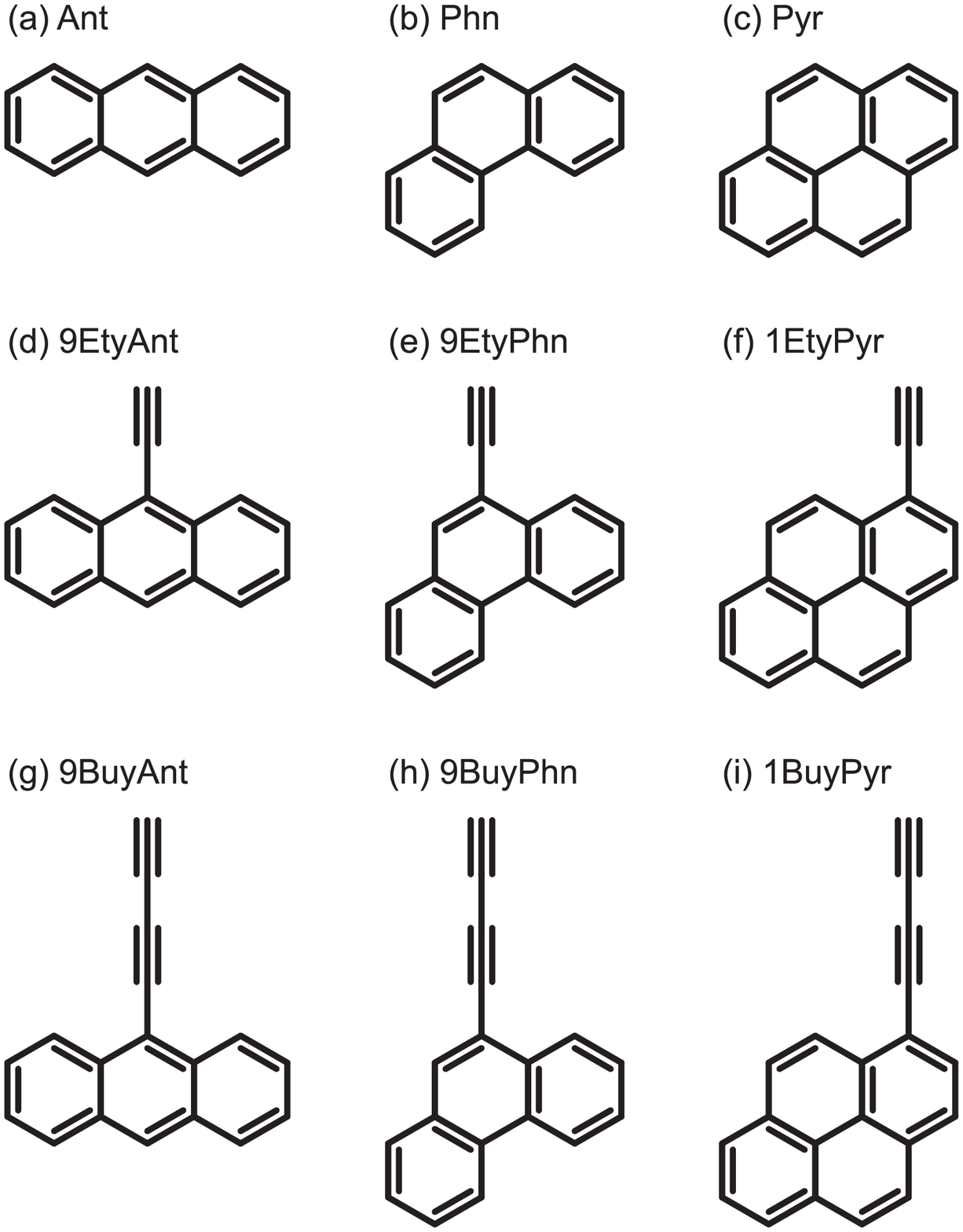}
\caption{Structures of the molecules presently studied: anthracene C$_{14}$H$_{10}$ (a), phenanthrene C$_{14}$H$_{10}$ (b), pyrene C$_{16}$H$_{10}$ (c), 9-ethynylanthracene  C$_{16}$H$_{10}$ (d), 9-ethynylphenanthrene C$_{16}$H$_{10}$ (e), 1-ethynylpyrene C$_{18}$H$_{10}$ (f), 9-butadiynylanthracene C$_{18}$H$_{10}$ (g), 9-butadiynylphenanthrene C$_{18}$H$_{10}$ (h), and 1-butadiynylpyrene C$_{20}$H$_{10}$ (i).\label{fig1}}
\end{figure}

Figure~\ref{fig1} shows the structures of the molecules studied. Alongside the non-substituted PAHs anthracene, phenanthrene, and pyrene, they include a selection of derivatives obtained by replacing an H atom with either an ethynyl ($-$C$_2$H) or a butadiynyl ($-$C$_4$H) group. Several of these substances were obtained from commercial providers. Anthracene (Aldrich, purity 97{\%}), phenanthrene (Aldrich, purity 98{\%}), 9-ethynylphenanthrene (Aldrich, purity 97{\%}), pyrene (Aldrich, purity 99{\%}), and 1-ethynylpyrene (ABCR, purity 96{\%}) were used as received.

The other substances, namely 9-ethynylanthracene, 9-butadiynylanthracene, 9-butadiynylphenanthrene, and 1-butadiynylpyrene, were synthesized in-house. They were prepared via Negishi coupling \citep{Negishi03} of the corresponding bromoarenes with an appropriate silylated alkyne followed by protodesilylation. The structure of these molecules was verified through NMR measurements. Moreover, the purity of the in-house samples was controlled by HPLC. The chromatograms yielded purities of 96, 93.5, $>$99, and $>$99{\%} for 9-ethynylanthracene, 9-butadiynylanthracene, 9-butadiynylphenanthrene, and 1-butadiynylpyrene, respectively. These four substances were kept under Ar atmosphere in vials stored in a Dewar flask filled with CO$_2$ ice and were used as synthesized.

For convenience, the names of the substances are abbreviated. Anthracene, phenanthrene, and pyrene are referred to as Ant, Phn, and Pyr, respectively. Each derivative is indicated with a prefix comprising the number that identifies the carbon atom carrying the side chain followed by Ety for ethynyl or Buy for butadiynyl. For instance, 9EtyAnt stands for 9-ethynylanthracene.

\subsection{Infrared spectroscopy} \label{ex:CsI}

The IR absorption spectra of the six polyynyl-substituted PAHs were measured from 2.5 to 67~$\mu$m at room temperature on grains embedded in CsI pellets. Each pellet was prepared by pressing 200 to 400~mg of a mixture of the sample with CsI powder. In every mixture, the mass ratio of sample to CsI was 1:500. The IR spectra were obtained by transmission using a Fourier transform IR spectrometer (Bruker 113v) with a resolution of 2~cm$^{-1}$ and an averaging over 32 measurements. Spectra were measured in two steps as different detectors had to be used, one for the 150--660~cm$^{-1}$ range, another for the 400--4000~cm$^{-1}$ domain. A pellet of pure CsI was used to provide a reference spectrum for background subtraction. Further correction was necessary and was carried out by subtracting a polynomial function fitted to the background.

\subsection{Matrix isolation spectroscopy} \label{ex:MIS}

The matrix isolation spectroscopy (MIS) setup was described in previous studies \citep{Steglich10,Rouille11}. Briefly, it consists of a commercial spectrophotometer (JASCO V-670 EX) coupled to a vacuum chamber by means of optical fibers. The chamber is equipped with a closed-cycle He cryocooler (Advanced Research Systems Inc. DE-204SL).

We used Ne (Linde, purity 99.995{\%}) as the matrix material. In the course of each experiment, a CaF$_\mathrm{2}$ substrate was placed into the vacuum chamber and cooled to a temperature of 6 to 7.5~K by the action of the cryocooler. The rare gas was fed into the vacuum chamber at a flow rate of 5~sccm (standard cubic centimeter per minute), and it condensed on the cold substrate. In order to produce PAH-doped Ne matrices, the rare gas was simply seeded with PAH molecules. Under normal conditions, most PAHs, including those presently studied, are in the solid phase. Still, molecules are released into the gas phase and the associated vapor pressure is temperature-dependent. Thus, the sample under study was kept in a small reservoir connected to the rare-gas line and its vapor pressure was optimized by varying its temperature.

So as to obtain the spectra of Ant, 9EtyAnt, and 9BuyAnt, the samples were kept at 18--24, 5.0--8.5, and 63~$\degr$C, respectively. The deposition of the respective matrices took 24, 105, and 95~min. For Phn, 9EtyPhn, and 9BuyPhn, temperatures of 20.5, 13.5, and 61~$\degr$C were used, respectively, and the corresponding matrices were formed in 3, 60, and 115~min. Note that, as an exception, the flow of Ne in the case of Phn was 7~sccm. Regarding Pyr, the sample was left at room temperature (21~$\degr$C) and the deposition of the matrix took 45~min. The temperature applied to 1EtyPyr was 40~$\degr$C and the matrix was grown in 50~min.

Attempts to obtain spectra of 1BuyPyr by following the procedure described above were unsuccessful. Whereas the heated 9BuyAnt and 9BuyPhn samples released enough molecules to dope Ne matrices, the 1BuyPyr powder did not, as inferred from the absence of 1BuyPyr bands from the measured spectra. Temperatures as high as 105~$\degr$C had been applied. It did not seem reasonable to consider higher temperatures in view of the behavior of the other two butadiynyl-substituted species compared to those of the corresponding ethynyl-substituted and parent PAHs. For this reason, we applied laser vaporization to 1BuyPyr. A little amount ($\sim$1~mg) of 1BuyPyr powder was spread over the surface of a CaF$_2$ substrate. In order to make the 1BuyPyr grains stick to the substrate, drops of methanol, one at a time, were deposited on them and let to evaporate. In the MIS chamber, the sample-carrying substrate was placed with its back to the Ne gas inlet and its sample-covered side facing the subtrate on which the matrix would form. Vaporization of 1BuyPyr was then obtained by directing a pulsed laser beam toward the sample. The laser source (Continuum Minilite II) emitted photons with a wavelength of 532~nm, in pulses of $\sim$10~ns duration and 1~mJ energy. It was operated with a repetition rate of 10~Hz. The laser spot at the surface of the sample had an approximate diameter of 2.7~mm, giving a fluence of 18~mJ~cm$^{-2}$. As the Ne flow was 5~sccm, the sample was exposed to the laser beam for a total duration of 20~min. Every 2~min, the position of the laser spot on the sample was shifted. Due to the limited size of the sample, some overlap of the irradiated spots could not be avoided. The sample darkened and blackened during the experiment, especially where the irradiated spots overlapped, and was not entirely vaporized.

When the bands of the $S_1 \leftarrow S_0$ transition were too weak for being clearly observed in the extended spectra, new measurements were carried out after further growth of the matrix, with an increased temperature of the oven when it was necessary.

All our measurements were carried out in steps of 2~{\AA}, with a scanning speed of 110~{\AA}~min$^{-1}$, and at a resolution of 2~{\AA}. The spectrophotometer was calibrated with an accuracy of 3~{\AA}. The spectra presented result from the averaging of up to nine measurements.

\subsection{Theoretical calculations} \label{ex:theory}

Theoretical calculations were carried out to assist in the interpretation of the various spectra. We have determined the theoretical structures of the molecules in their electronic ground state. For this purpose, calculations based on the density functional theory (DFT) were performed with the Gaussian 09 software \citep{Gaussian09}. The B3LYP functional was used in conjunction with the cc-pVTZ basis set. The tight convergence criteria implemented in the software were applied when optimizing the molecular geometries. Optimization procedures were conducted while forcing Ant and Pyr into $D_{2h}$ geometry, Phn, 9EtyAnt, and 9BuyAnt into $C_{2v}$, and the other molecules into $C_s$. Vibrational frequencies were computed to verify that the optimized structures corresponded to minima in the potential surfaces. In addition, the results included the values of the permanent electric dipole moment. The predefined ultrafine grid of Gaussian 09 was used in all DFT-based calculations.

Electronic states and transitions were calculated at the geometries determined for the electronic ground states. Both the ZINDO semi-empirical model and the time-dependent density functional theory (TD-DFT) approach were used. The TD-DFT calculations were carried out with the B3LYP functional and the cc-pVTZ basis set already employed for the geometry optimization of the electronic ground states. The ultrafine grid was used in these calculations as well.

\section{Results}

To our knowledge, the UV/vis spectra of the butadiynyl-substituted PAHs are reported here for the first time ever, independently of the experimental conditions. While the spectra of the ethynyl-substituted PAHs were measured before in solutions at room temperature \citep{Marsh00}, our low-temperature, matrix-isolated spectra are the first to be published. Finally, the full UV/vis spectrum of Ant isolated in a Ne matrix is also presented for the first time. The spectra we obtained for Phn and Pyr are in agreement with previous measurements \citep{Salama94,Salama95}.

The IR spectra of the butadiynyl-substituted PAHs are also presented for the first time. Our spectra of 9EtyAnt and 9EtyPhn are in very good agreement with results obtained from films condensed on substrates kept at 77~K \citep{Visser98}. We have not found detailed reports on the IR spectrum of 1EtyPyr, preventing comparison with earlier results. The IR spectra of the non-substituted PAHs, measured under various conditions, can be found in the literature, e.g., in \citet{Hudgins98} for PAHs isolated in Ar matrix and in \citet{Karcher85} for measurements in KBr pellets and in solutions.

\subsection{UV spectra} \label{res:UV}

\subsubsection{Anthracene and its derivatives} \label{res:UV:Ant}

\begin{figure}
\epsscale{0.50}
\plotone{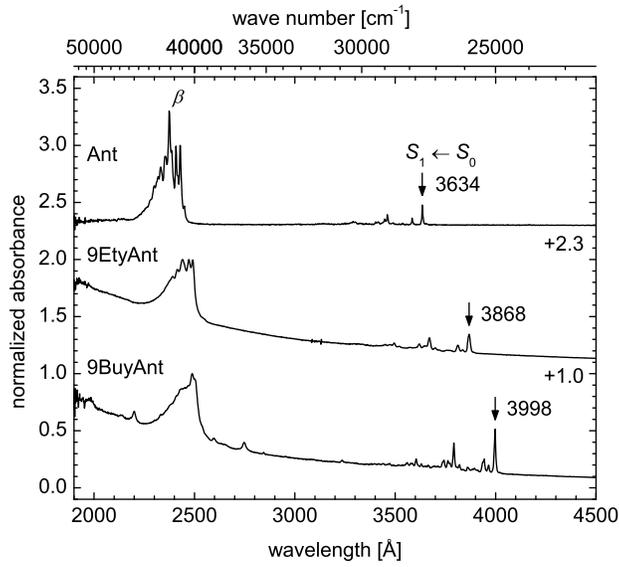}
\caption{Absorption spectra of Ant, 9EtyAnt, and 9BuyAnt isolated in Ne matrices at 6~K. The origin band of each $S_1 \leftarrow S_0$ transition is indicated by an arrow and labelled with its peak position. Each spectrum has been normalized with respect to the intensity of its highest peak. For the purpose of better comparison, the upper two spectra were shifted upward by the amounts indicated at the right-hand side. In the spectra of 9EtyAnt and 9BuyAnt, the background that increases toward shorter wavelengths is due to a strong scattering effect in the matrix.\label{fig2}}
\end{figure}

Figure~\ref{fig2} shows the electronic spectra of Ant and its derivatives. In the spectrum of Ant, the features of the well-known $S_1 \leftarrow S_0$ transition are clearly seen. They consist of the origin band at 3634~{\AA} and the associated vibronic bands that form a progression on its blue side to approximatly 3000~{\AA}. At shorter wavelength, a broad feature dominates the spectrum, exhibiting multiple peaks with a maximum at 2374~{\AA}. In accordance with Clar's notation \citep{Clar50}, it is denoted by $\beta$-band.

The spectrum of 9EtyAnt is similar in its general aspect. It shows a series of bands with its origin at 3868~{\AA}, which we assign to the $S_1 \leftarrow S_0$ transition. One can note that the band pattern differs slightly from the one given by the $S_1 \leftarrow S_0$ transition of Ant, due to the presence of a new band next to the origin. The frequency of the corresponding vibrational mode is $\sim$209~cm$^{-1}$. With 9EtyAnt too, the spectrum is dominated by a broad feature attributed to the $\beta$-band, which reaches its maximum at 2440~{\AA}. In comparison with the spectrum of Ant, the bands are redshifted, by 234~{\AA} in the case of the origin of the $S_1 \leftarrow S_0$ transition. The bands are also broadened, and the peaks surmounting the $\beta$-band are less separated. Finally, the ratio of the peak intensity of the origin band of the $S_1 \leftarrow S_0$ transition to that of the $\beta$-band is higher.

In the case of 9BuyAnt, one observes again a spectrum similar to that of Ant. It is affected by the differences already noted in the spectrum of the ethynyl-substituted molecule, only more strongly. The redshifts are larger as the origin band of the $S_1 \leftarrow S_0$ transition is found at 3998~{\AA}, that is 364~{\AA} more to the red than in the spectrum of Ant, and the $\beta$-band arises at approximately 2500~{\AA}. A single maximum surmounts the $\beta$-band suggesting a larger broadening than in the spectrum of 9EtyAnt. The ratio of the peak intensities of the origin band of the $S_1 \leftarrow S_0$ transition relative to the $\beta$-band has further increased. A new band observed next to the origin of the $S_1 \leftarrow S_0$ transition of 9EtyAnt is also present in the spectrum of 9BuyAnt, indicating a vibrational mode with a frequency of $\sim$216~cm$^{-1}$. Other minor peaks are detected, notably at 2200, 2746, and 3236~{\AA}, which belong neither to the $S_1 \leftarrow S_0$ transition nor to the $\beta$-band.

The vibrational mode with a frequency of $\sim$210~cm$^{-1}$ that gives rise to a band in the $S_1 \leftarrow S_0$ transition of 9EtyAnt and 9BuyAnt, but not in that of Ant, is an in-plane CCC bending motion. Selection rules forbid the excitation of this mode in the $S_1 \leftarrow S_0$ transition of Ant, which is a molecule with $D_{2h}$ symmetry. The rules change for 9EtyAnt and 9BuyAnt as the molecular symmetry becomes $C_{2v}$ upon the addition of the side chains. According to the IR spectra introduced in Section~\ref{res:IR}, the mode has a frequency of 232 and $\sim$220~cm$^{-1}$ in the ground states of 9EtyAnt and 9BuyAnt, respectively, as measured on grains dispersed in CsI pellets. We found unscaled (scaled) theoretical values of 234 (228) and 218 (212)~cm$^{-1}$, respectively, with the calculations presented in Sections~\ref{ex:theory} and \ref{res:theo}.

\subsubsection{Phenanthrene and its derivatives} \label{res:UV:Phn}

\begin{figure}
\epsscale{0.50}
\plotone{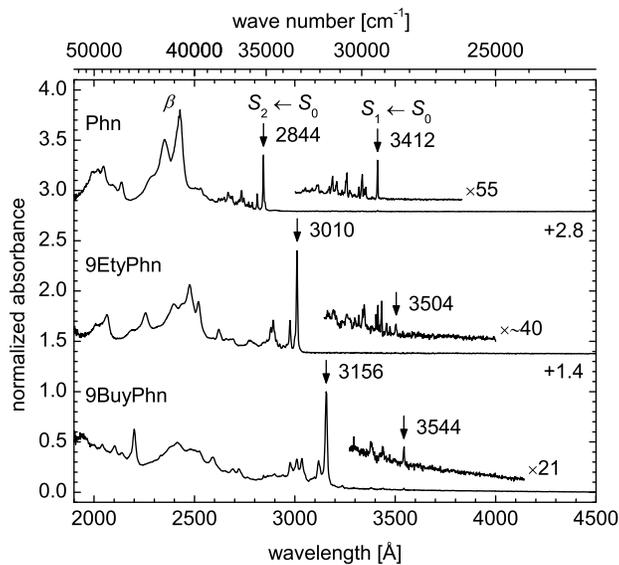}
\caption{Absorption spectra of Phn, 9EtyPhn, and 9BuyPhn isolated in Ne matrices at 6~K. Arrows labelled with wavelengths indicate the peak positions of the origin bands of the $S_1 \leftarrow S_0$ and $S_2 \leftarrow S_0$ transitions. The vertical expansion factors given for the $S_1 \leftarrow S_0$ spectra have been determined by various means and provide merely an order of magnitude. Other details of presentation are explained in the caption of Figure~\ref{fig2}.\label{fig3}}
\end{figure}

The electronic spectra of Phn and its derivatives are displayed in Figure~\ref{fig3}. In the spectrum of Phn, the $S_1 \leftarrow S_0$ transition gives extremely weak features, with the origin band arising at 3412~{\AA}. At shorter wavelength, one finds the strong bands of the $S_2 \leftarrow S_0$ transition, which has its origin at 2844~{\AA}. Then one finds the features of the $\beta$-band, the strongest of which dominates the whole spectrum and peaks at 2428~{\AA}. It is attributed to the origin of this electronic transition. Taking into account the accuracy of the measurements, the positions we report are in excellent agreement with those of \citet{Salama94}, which were 3411, 2834, and 2430~{\AA}, respectively.

Regarding 9EtyPhn, one finds a spectrum somewhat similar to that of Phn, with the obvious difference that the $\beta$-band no longer dominates the spectrum. The main peak in the region of the $\beta$-band, at 2476~{\AA}, is redshifted by 48~{\AA} in comparison with the $\beta$-band of Phn. Attributing this peak to the origin of the $\beta$-band is tentative as several maxima arise around this position. They most likely belong to several electronic transitions as they cannot be easily interpreted as a unique vibrational progression. The dominating feature in the spectrum of 9EtyPhn corresponds to the origin band of the $S_2 \leftarrow S_0$ transition, which is found at 3010~{\AA}, 166~{\AA} more to the red than the same transition in Phn. The vibronic pattern is very similar to what is observed for Phn, only the bands are slightly broadened. At longer wavelength, a series of weak bands is attributed to the $S_1 \leftarrow S_0$ transition. We assign the band with the longest wavelength, at 3504~{\AA}, to the origin of the transition. The vibrational pattern is complex, with a weak origin band, which is different from what is observed in Phn.

The UV/vis spectrum of 9BuyPhn resembles that of 9EtyPhn. In the region where the $\beta$-band is expected, however, one finds a very broad feature, for which it is not possible to determine the origin wavelength. Due to this broadening, in comparison with the origin band of the $S_2 \leftarrow S_0$ transition, the $\beta$-band has an even lower strength than in the spectrum of 9EtyPhn. The origin band of the $S_2 \leftarrow S_0$ transition dominates the spectrum at 3156~{\AA}. Like the other bands of the same transition, it is broader and redshifted by 146~{\AA} compared to the corresponding features in the 9EtyPhn spectrum. As for the $S_1 \leftarrow S_0$ transition, its spectrum is very weak and consists of a very small number of features. We assign the band at 3544~{\AA} to the origin of the transition. One notes that the bands of the $S_1 \leftarrow S_0$ transition are roughly twice as intense with respect to the $S_2 \leftarrow S_0$ bands in the spectra of Phn and 9EtyPhn.

\subsubsection{Pyrene and its derivatives} \label{res:UV:Pyr}

\begin{figure}
\epsscale{0.50}
\plotone{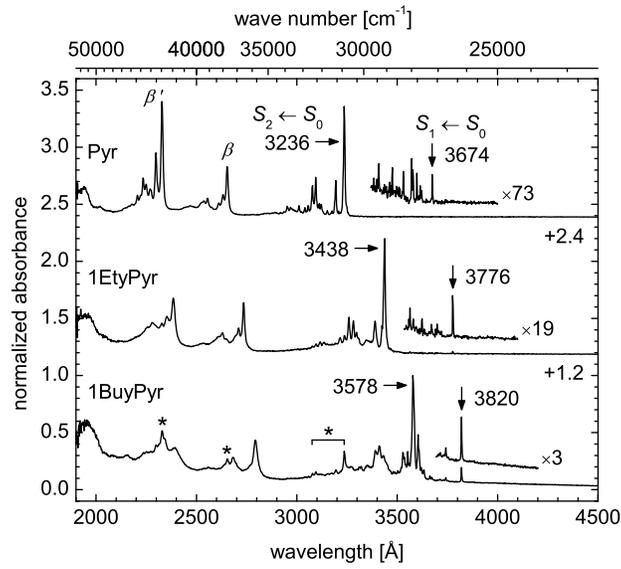}
\caption{Absorption spectra of Pyr, 1EtyPyr, and 1BuyPyr isolated in Ne matrices at 6~K. Arrows labelled with wavelengths indicate the peak positions of the origin bands of the $S_1 \leftarrow S_0$ and $S_2 \leftarrow S_0$ transitions. Asterisks and a horizontal bracket in the spectrum of 1BuyPyr mark peaks that are essentially due to Pyr. Laser vaporization was used for 1BuyPyr. The vertical expansion factors given for the $S_1 \leftarrow S_0$ spectra have been determined by various means and provide merely an order of magnitude. Other details of presentation are explained in the caption of Figure~\ref{fig2}.\label{fig4}}
\end{figure}

Figure~\ref{fig4} depicts the electronic spectra of Pyr and its derivatives. The spectrum of Pyr is dominated by the features of three systems, which are the $S_2 \leftarrow S_0$ transition, the $\beta$-band, and the $\beta'$-band. They give rise to maxima at 3236, 2652, and 2328~{\AA}, respectively, which mark the origins of the systems. These values are in agreement with those obtained by \citet{Salama95} from a less resolved spectrum. We found the origin band of the very weak $S_1 \leftarrow S_0$ transition at 3674~{\AA}.

The spectrum of 1EtyPyr is similar to that of Pyr with broadened and redshifted transitions. In this instance, however, the spectrum is clearly dominated by the peak that marks the origin of the $S_2 \leftarrow S_0$ transition at 3438~{\AA}. The origins of the $\beta$- and $\beta'$-bands are found at 2736 and 2384~{\AA}, respectively, while the origin of the $S_1 \leftarrow S_0$ transition is at 3776~{\AA}. One also notes, in comparison with the spectrum of the parent PAH, that the relative peak intensity of the origin band of the $S_1 \leftarrow S_0$ transition is higher.

Finally, in the spectrum of 1BuyPyr, the origins of the $S_2 \leftarrow S_0$ transition, $\beta$-band, and $\beta'$-band are found at 3578, 2794, and (tentatively) 2394~{\AA}, respectively. In comparison with the spectrum of 1EtyPyr, the features are further broadened and redshifted. The origin band of the $S_1 \leftarrow S_0$ transition is found at 3820~{\AA}. Its relative peak intensity has become much stronger.

One can note in the spectrum of 1BuyPyr that the origin of the $S_2 \leftarrow S_0$ transition arises as a complex vibronic structure. A matrix site effect or phonon side wings can be ruled out as they would affect the origin band of the $S_1 \leftarrow S_0$ transition in the same way. We propose that the structure is caused by an interaction between the $S_2$ state and the vibrational manifold of the $S_1$ state. Studies in molecular beams have shown that this interaction takes place in Pyr (see \citealt{Rouille04} and references therein). The resulting vibronic structure in the $S_2 \leftarrow S_0$ transition is not noticeable in matrix-isolated spectra of Pyr because the matrix-induced band broadening makes the bands overlap. Nothing forbids the interaction to occur in 1EtyPyr and 1BuyPyr. In the case of 1BuyPyr, the difference between the energies of the $S_2$ and $S_1$ states is small. It is approximately 1770~cm$^{-1}$, as compared with 3684 and 2604~cm$^{-1}$ for Pyr and 1EtyPyr, respectively. Consequently, the interaction is strong and it involves the vibrational manifold of the $S_1$ state where fundamental vibrational levels are found. Thus, in the vicinity of the origin of the $S_2 \leftarrow S_0$ transition, it gives rise to a strong and discrete vibronic structure. The increased strength of the origin band of the $S_1 \leftarrow S_0$ transition is also an effect of this interaction.

Considering all UV/vis spectra of Figures~\ref{fig2}--\ref{fig4}, the lower-energy transitions rather resemble those of the parent PAH molecules in terms of vibrational pattern and bandwidth. In contrast, the higher-energy transitions are strongly broadened, causing the lowering of their peak intensities. As this broadening increases with the length of the side group, we relate it to the number of its vibrational modes. Thus, the broadening would correspond to fast deexcitation through internal conversion and not to a matrix effect.

\subsection{IR spectra} \label{res:IR}

\begin{figure}
\epsscale{1.00}
\plotone{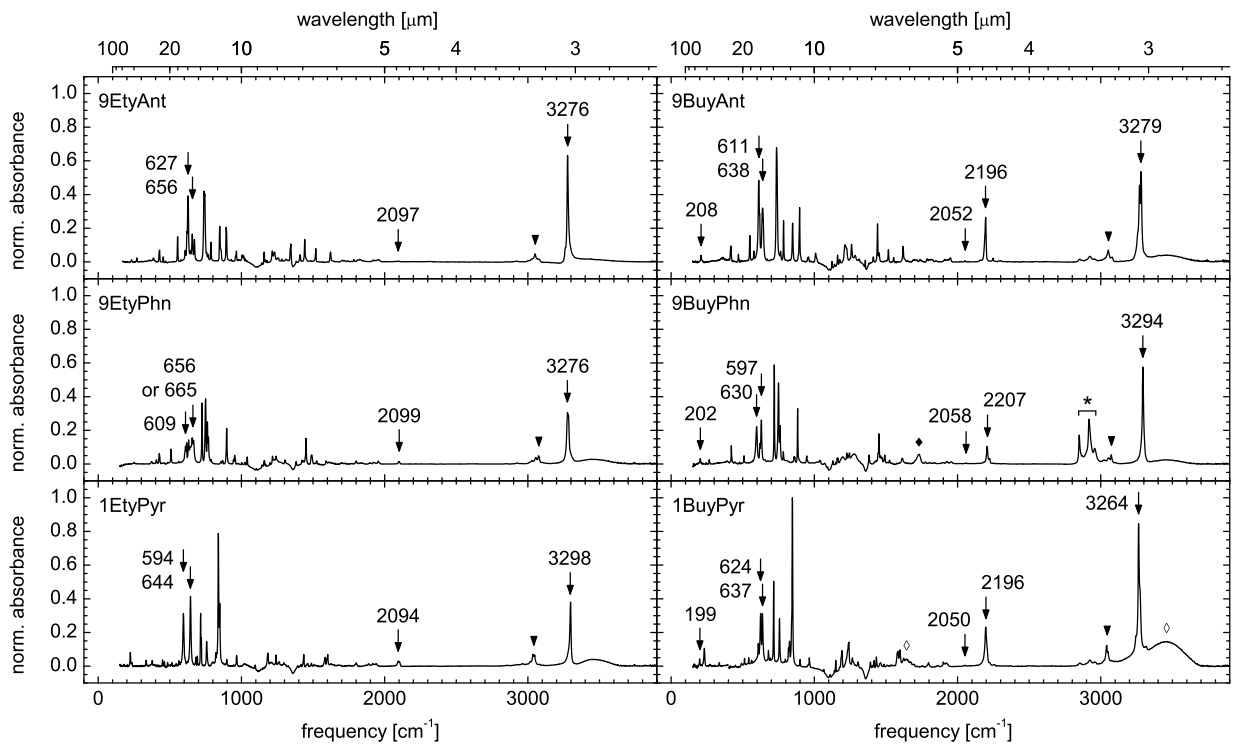}
\caption{Normalized IR absorption spectra of the polyynyl-substituted PAHs in CsI pellets. Arrows labelled with frequencies indicate bands due to remarkable vibrational modes of the side chains. These are the in-plane waving of the butadiynyl side chain near 200~cm$^{-1}$, the C$\equiv$C$-$H out-of-plane and in-plane bending modes in the 580--680~cm$^{-1}$ interval, the C$\equiv$C stretching mode at $\sim$2095~cm$^{-1}$ for the unique mode of the ethynyl group and at $\sim$2050 and $\sim$2200~cm$^{-1}$ for the antisymmetric and symmetric modes of the butadiynyl group, respectively. In addition, the acetylenic CH stretching mode falls between 3260 and 3300~cm$^{-1}$. Full triangles mark the aromatic CH stretching modes. The full diamond and the asterisk in the spectrum of 9BuyPhn indicate bands attributed to impurities. The band labelled with a full diamond is characteristic of a C$=$O stretching mode while the three peaks denoted by the asterisk correspond to CH stretching vibrations in groups where the C atom is $sp^3$-hybridized. The latter features are present in the other spectra, though with weaker intensities. In the spectrum of 1BuyPyr, the open diamonds mark broad features due to water, which can also be seen in the other panels with various strengths. In all panels, the dips at $\sim$1100 and $\sim$1360~cm$^{-1}$ are caused by the Christiansen effect in the reference CsI pellet.\label{fig5}}
\end{figure}

Figure~\ref{fig5} shows the IR absorption spectra of the polyynyl-substituted PAHs. The spectra have been first normalized with respect to the mass of sample powder in each pellet and the molecular weight of each species, and then with respect to the strongest band.

Some of the bands characteristic of the PAH family are not significantly affected by the substitution of a H atom with a much heavier polyynyl chain. Thus, the nonresolved group of weak bands at $\sim$3050~cm$^{-1}$ (3.28~$\mu$m) corresponds to the aromatic CH stretching modes. In spite of the substitution, their frequencies are similar to those observed for the CH stretching modes of the parent PAHs \citep{Hudgins98}. Another feature characteristic of PAHs is a strong band, the strongest or second strongest in our spectra, which is found between 700 and 900~cm$^{-1}$ (14.3 and 11.1~$\mu$m). It is due to the CH bending mode in which all H atoms move out-of-plane in phase. While the position of the band varies depending on the PAH structure, it is not much affected by the length of the side group.

In contrast, the substitution has a dramatic effect on other modes as it modifies the symmetry of the molecules. For instance, Ant has an out-of-plane aromatic CH bending mode with a frequency of $\sim$824~cm$^{-1}$ (value of the theoretical harmonic frequency scaled by a factor of 0.97), which is inactive with respect to IR absorption because it conserves the inversion center of the molecule. After substitution of one of the H atoms on the central ring for a much heavier side group, this mode becomes active, giving rise to a band of medium intensity at 849~cm$^{-1}$ in the spectra of 9EtyAnt and 9BuyAnt.

Other bands, which are indicated with arrows in Figure~\ref{fig5}, are specific to polyynyl side groups and correlate with the vibrational frequencies in the free molecules, ethyne and butadiyne \citep{Shimanouchi72}. Accordingly, all spectra show more or less clearly a pair of bands of medium strength in the 625~cm$^{-1}$ (16~$\mu$m) region. They correspond, in the order of increasing frequency, to the C$\equiv$C$-$H out-of-plane and in-plane bending modes, which are shared by all polyynyl side groups. All spectra also display a band of high intensity in the 3260--3300~cm$^{-1}$ (3.030--3.068~$\mu$m) interval which is caused by the acetylenic CH stretching vibration. Except for 9EtyAnt and 9BuyAnt, the band is stronger for the butadiynyl-substituted PAHs than for the ethynyl-substituted ones. As the band has a similar intensity in the spectra of 9BuyAnt, 9BuyPhn, and 1BuyPyr, the normalization of the spectrum of 9EtyAnt may be incorrect due to an inaccurate estimation of the mass of this sample.

On the other hand, ethynyl- and butadiynyl-substituted PAHs can be distinguished by examining the bands arising from the acetylenic CC stretching modes. The ethynyl group possesses a single C$\equiv$C bond and, as a consequence, a single stretching mode, which causes a very weak band at $\sim$2095~cm$^{-1}$ (4.77~$\mu$m). With two C$\equiv$C bonds, the butadiynyl group has two stretching modes, antisymmetric and symmetric, that cause bands at $\sim$2050 and $\sim$2200~cm$^{-1}$ (4.88 and 4.55~$\mu$m), respectively. The band due to the antisymmetric stretching is very weak while the other has medium intensity. Another means to distinguish between ethynyl- and butadiynyl-substituted PAHs is to look for the band arising from the in-plane waving mode of the butadiynyl chain near 200~cm$^{-1}$ (50~$\mu$m). Although this band is weak, it can be clearly seen in our spectra. According to calculations, the band caused by the corresponding out-of-plane motion would be weaker, in various proportions, and would lie between 120 and 160~cm$^{-1}$ (62 and 83~$\mu$m), at the limit of the range covered by our measurements. Therefore, it cannot be found in our spectra.

\subsection{Theory} \label{res:theo}

In Sections~\ref{res:UV} and \ref{res:IR}, the measured spectra have been interpreted with the assistance of the results of theoretical calculations and with the knowledge obtained from the study of non-substituted PAH molecules. Our interpretation of the laboratory IR spectra takes into account the theoretical IR spectra displayed in Figure~\ref{fig6}. These spectra reproduce the measurements well enough to allow us to assign the experimental IR bands. Still, as they were obtained with DFT-based calculations, they differ from the laboratory spectra in ways that were already reported for the case of other PAHs (see \citealt{Rouille11} and references therein). First, the theoretical frequencies are larger than the observed frequencies. This is not unexpected since the theoretical values do not take anharmonicity into account. It is common to correct the theoretical frequencies by applying a scaling factor, if necessary by taking the aromatic CH stretching frequencies into account separately. Detailed discussions of this issue, first addressed by \citet{Langhoff96}, and DFT spectra for many PAHs can be found in \citet{Malloci07} and \citet{Bauschlicher10}. In this study, a value of 0.959 is found by fitting the scaling factor as the set of data includes simultaneously frequencies of the six polyynyl-substituted PAHs. When the bands are divided in two groups, those at frequencies below 2000~cm$^{-1}$ and those at frequencies above 2000~cm$^{-1}$, scaling factors of 0.974 and 0.953 are obtained for the low- and high-frequency domains, respectively. The theoretical and measured spectra also differ in their intensity patterns as the relative intensities of the aromatic CH stretching bands are much higher in the theoretical spectra than in the measured spectra (see \citealt{Rouille11} and references therein). We find that this effect applies to the acetylenic CC and CH stretching bands as well.

Another property of the molecules that has been determined is their permanent electric dipole moment in the electronic ground state. Table~\ref{table1} gives the theoretical values for the parent PAHs and their derivatives. Among the parent PAHs, only Phn possesses a permanent electric dipole moment as the molecule does not have an inversion center. Its value, however, is extremely small and has not been determined experimentally to our knowledge. A significant dipole moment appears when an ethynyl side group is present. For comparison, the theoretical values we have obtained for 9EtyAnt (0.5893~D), 9EtyPhn (0.7466~D), and 1EtyPyr (0.9251~D) are close to the value measured for ethynylbenzene, that is 0.656 $\pm$ 0.005~D \citep{Cox75}. Our results show that the electric dipole moment is larger with a butadiynyl side group than with an ethynyl side group by a factor of two approximately.

Electronic states and vertical transition energies are given in Table~\ref{table2}. While theoretical vibrational frequencies are calculated with an accuracy of a few cm$^{-1}$, the uncertainty on theoretical electronic transition energies is far greater, i.e., of the order of 0.5~eV for the values computed with the TD-DFT approach \citep{Dierksen04}. In the case of PAH molecules, the theoretical determination of electronic transitions is usually challenging, because of the close proximity of the electronic states. Interactions, which are not taken into account in the usual models, may take place between them. As noticed in the UV/vis spectra of the phenanthrenes and pyrenes, the energy spacing between the first two excited states decreases upon substitution with the ethynyl group and decreases further with the butadiynyl group. The calculation of the electronic states and vertical transition energies gives qualitatively reasonable results for the anthracenes, even though quantitative differences exist between the results obtained with the ZINDO and TD-DFT approaches. This is also true for Phn and 9EtyPhn. For 9BuyPhn and the pyrenes, the calculation fails to give the correct order for the two electronic states with the lowest energies (with the exception of the ZINDO model applied to Pyr).

We have examined the molecular orbitals calculated with the ZINDO model and also those obtained with the TD-DFT approach. The electronic transitions we have observed are attributed to $\pi$-$\pi^*$ transitions. In every substituted species we have studied, the PAH moiety and the side group contribute together to a number of $\pi$ orbitals, notably the highest occupied and lowest unoccupied molecular orbitals (HOMO and LUMO). For example, the HOMO and the LUMO of 9BuyAnt are depicted in Figure~\ref{fig7}, along with the next orbitals, i.e., the HOMO$-$1 and the LUMO$+$1. While the latter two orbitals do not comprise contributions from the side group in the substituted anthracenes and phenanthrenes, they do in the substituted pyrenes.

Because electronic transitions in substituted PAHs involve molecular orbitals with contributions from both the PAH moiety and the side group, more electrons are involved in the transitions of a substituted PAH compared to the parent PAH. Thus, larger oscillator strengths can be expected for the substituted molecules. Such a trend is observed in Table~\ref{table2}, except for the $\beta$-band of the anthracenes, for which the oscillator strength decreases as the length of the side group increases. The growth of the oscillator strength remains however modest as it barely reaches an order of magnitude.

\section{Discussion}

\subsection{Relevance of the laboratory spectra} \label{disc:labor}

Ideally, the interpretation of observational (astrophysical) spectra makes use of reference measurements obtained in the laboratory under the conditions of the astrophysical object. In general, it is not possible to exactly reproduce these conditions. Therefore, either they are mimicked or information is extracted from laboratory measurements obtained under various conditions.

Although interstellar PAHs have been revealed by their collective IR emission features, the AIBs, the IR spectroscopy of PAHs is most often performed in the laboratory in absorption, either at cryogenic temperature on molecules isolated in a rare gas matrix (such as in MIS) or, more simply, at room temperature on grains and crystallites embedded in salt pellets. As we have opted for the latter technique, the experimental conditions do not allow us to compare directly our measurements with the AIBs, which are attributed to populations of free-flying PAH molecules, nor do they allow us to compare them with the absorption bands caused by other free-flying molecules such as benzene, the only PAH-like species identified in space so far \citep{Cernicharo01b}. Indeed, due to the interaction of the PAH molecules with their neighbors, the IR absorption bands measured in salt pellets are affected by shifts and variations of the relative intensities in comparison with the gas-phase absorption spectra. Moreover, the absorption bands presently measured in CsI pellets with a full width at half maximum (FWHM) of $\sim$5--6~cm$^{-1}$ are narrower than the emission bands, which are expected to show a FWHM of $\sim$30~cm$^{-1}$ \citep{Allamandola99b}. Finally, the band intensities observed in absorption and emission may differ as, in the latter case, they depend on the energy distribution among the vibrationally excited modes in addition to the transition moment.

After they were found in IR sources, PAH molecules were proposed as candidates for the carriers of the diffuse interstellar bands (DIBs), which are interstellar absorption features observed between 4000 and 18\,000~{\AA} against background stars (see \citealt{Geballe11} and references therein). The UV/vis absorption spectra of PAHs can be measured more easily under conditions that mimic those of the ISM, e.g., by applying cavity ring-down laser absorption spectroscopy on jet-cooled molecules \citep{Motylewski00,Gredel11,Salama11}. Such spectra can be directly compared with the DIBs. The drawback of laser-based spectroscopy techniques is the usually narrow wavelength range they can cover. As a consequence, they represent the ultimate step in testing candidates for the carriers of the DIBs. For a first approach to the spectral properties of a new species, it is preferable to apply MIS since measurements can cover very large wavelength ranges in a single scan while using little sample. The UV/vis absorption spectra we present are measured on PAH molecules isolated in Ne matrices at 6--7.5~K. Compared to spectra measured under conditions relevant to the ISM, the bands are expected to be redshifted by an amount approximately proportional to the variation of the polarizability of the molecule upon the corresponding transition \citep[see, e.g.,][]{Gredel11}. Due to the low polarizability of Ne, the matrix-induced shifts are small, of the order of 1$\%$ or less. Therefore, matrix-isolated spectra represent valuable data because they reveal the general aspect of the spectra, which can be sufficient to demonstrate that a molecule cannot be a DIB carrier.

\subsection{Polyynyl-substituted PAHs and astrophysics} \label{disc:astro}

\subsubsection{Formation} \label{disc:astro:form}

The present study was motivated by the conjecture that PAHs carrying $sp$-bonded chains may be present in IR emission regions. It was put forward by \citet{Duley09} following their observation of both aromatic structures and $sp$-bonded chains in laboratory analogs of cosmic carbonaceous grains. These analogs were produced by laser ablation of graphite and subsequent condensation of the ablated carbon atoms in the gas phase \citep{Duley09}. Another study on carbonaceous grains synthesized in a similar fashion not only revealed aromatic structures and $sp$-bonded chains, it also disclosed acetylenic CH bonds, though in a very low amount compared to the $sp$-bonded chains \citep{Jaeger08}. The lack of acetylenic CH bonds in the former case and its very small amount in the latter indicated that the chains were essentially bridges between aromatic structures \citep{Duley09} or fullerene fragments \citep{Jaeger08,Jaeger09} rather than polyynyl side groups. Following laser ablation of graphite, carbonaceous grains form in the gas phase at high temperatures, in the range 4000--6000~K \citep{Jaeger08,Jaeger09}. They are considered as analogs of grains expected in the hot environment of carbon-rich stars or in the material ejected from supernovae \citep{Jaeger09}. We can conclude that the conditions in these environments are not favorable to the formation of polyynyl-substituted PAHs.

However, polyynyl-substituted PAHs would rather form at lower temperatures. \citet{Jaeger07} studied mixtures of molecules extracted from soots produced by IR laser-induced pyrolysis of ethylene, either pure or mixed with acetylene or benzene. Temperatures lower than 1800~K were found essential to the formation of PAH-rich soots \citep{Jaeger07,Jaeger09}. The absorption spectra of the mixtures extracted with chloroform (CHCl$_3$) showed bands at 2104 and 3288~cm$^{-1}$ (4.75 and 3.04~$\mu$m, respectively). Figure~\ref{fig8} reproduces the spectrum of one of the extracted mixtures \citep[sample CP85a of][]{Jaeger07} and compares it with the IR spectrum of 9EtyPhn from Figure~\ref{fig5}. The comparison leads us to attribute the bands at 4.75 and 3.04~$\mu$m to the stretching modes of C$\equiv$C bonds and acetylenic CH bonds, respectively. Their positions and relative intensities indicate the presence of ethynyl groups carried by PAHs. Since the formation of the molecules contained in the soot took place under conditions relevant to the environment of evolved stars \citep{Jaeger09}, e.g., asymptotic giant branch stars, it can be concluded that PAHs carrying ethynyl groups can form in such regions. Conclusions cannot be drawn concerning longer polyynyl side groups which were not detected in the soot extract.

The formation of ethynyl-substituted PAHs in the environment of stars can be explained by the H-abstraction-C$_2$H$_2$-addition (HACA) mechanism that contributes to the formation of larger PAHs \citep{Frenklach85,Frenklach89}. Indeed, ethynyl-substituted PAHs are intermediate products during the growth process \citep{Frenklach89,Wang94}. It can take place in the circumstellar envelopes of carbon-rich stars, in regions where the temperatures are in the range between 900 and 1100~K \citep{Frenklach89}.

Interestingly, polyynyl-substituted PAHs may form in the ISM itself, at temperatures much lower than those required by the HACA mechanism. In a combined theoretical and experimental study, \citet{Jones11} showed that benzene can form under conditions found in cold molecular clouds through the reaction of the ethynyl radical (C$_2$H) with 1,3-butadiene (CH$_2$CHCHCH$_2$). Starting with benzene, the so-called ethynyl addition mechanism (EAM) introduced by \citet{Mebel08} would lead to the formation of larger aromatic species \citep{Mebel08,Jones11}, first of all ethynylbenzene (also known as phenylacetylene) \citep{Goulay06,Jones10}. It would also lead to the formation of polysubstituted benzenes carrying ethynyl and butadiynyl groups, and, possibly, similarly polysubstituted PAHs. This mechanism is of great interest, for ethynyl is abundant in the ISM \citep{Tucker74} and in diffuse clouds \citep{Lucas00}. As the polyynyl radicals butadiynyl C$_4$H \citep{Guelin78}, hexatriynyl C$_6$H \citep{Suzuki86,Guelin87}, and octatetraynyl C$_8$H \citep{Cernicharo96} have been identified in space by microwave emission observations, various other products can be formed if they can also react with PAH molecules at low temperature. The relevance of the EAM and the other reactions just mentioned to the chemistry of the ISM needs to be examined especially with regard to timescales if more information on the corresponding reaction rates and reactant densities is available.

\subsubsection{Photostability} \label{disc:astro:stab}

Molecules formed in circumstellar envelopes or in diffuse molecular clouds are exposed to UV photon bombardment. Therefore the question of the photostability of the polyynyl-substituted PAHs arises. Data regarding this particular class of molecules are not available. On the other hand, laboratory studies on the photostability of other PAHs have been reported \citep{Jochims97,Jochims99}. \citet{Jochims99} found that methyl- and vinyl-substituted PAHs (H$_3$C$-$ and H$_2$C$=$CH$-$PAHs) have a lower photostability than their parent PAHs. This suggests that the photostability of a PAH molecule is weakened upon substitution with a hydrocarbon group. The extent of the effect, however, depends on the electronic structure of the substituent. It was observed that the effect was less pronounced upon substitution with a vinyl group than with a methyl group, probably on account of the CC double bond of the vinyl group. Similarly, the experiments revealed that the photostability of diphenylacetylene, which consists of two phenyl moieties connected by a $-$C$\equiv$C$-$ chain (H$_5$C$_6-$C$\equiv$C$-$C$_6$H$_5$), was higher than that of biphenyl, which consists of two phenyl moieties linked by a CC single bond (H$_5$C$_6-$C$_6$H$_5$) \citep{Jochims97,Jochims99}. The higher photostability was attributed to the presence of the CC triple bond, which increased the overall conjugation of the $\pi$ orbitals \citep{Jochims99}.

Thus, we can assume that polyynyl-substituted PAHs, like other substituted PAHs, have a lower photostability than their parent molecules. Taking into account how the conjugation of the $\pi$ orbitals affects the photostability, we can also assume a higher photostability of polyynyl-substituted PAHs compared to methyl-substituted PAHs and, by generalizing, PAHs carrying $sp^3$-bonded aliphatic chains. Precisely these latter species may exist in IR sources despite the local flow of UV photons. Indeed, an IR emission band at 3.4~$\mu$m has been attributed to the stretching vibration of an aliphatic C($sp^3$)H bond, which may be a component of side groups attached to PAH molecules \citep[see][]{Tielens08}. Thus, polyynyl-substituted PAHs, which have a higher photostability, may withstand the conditions that prevail in IR sources, and those in the ISM as well.

\subsubsection{Search for IR bands} \label{disc:astro:IR}

\begin{figure}
\epsscale{1.00}
\plotone{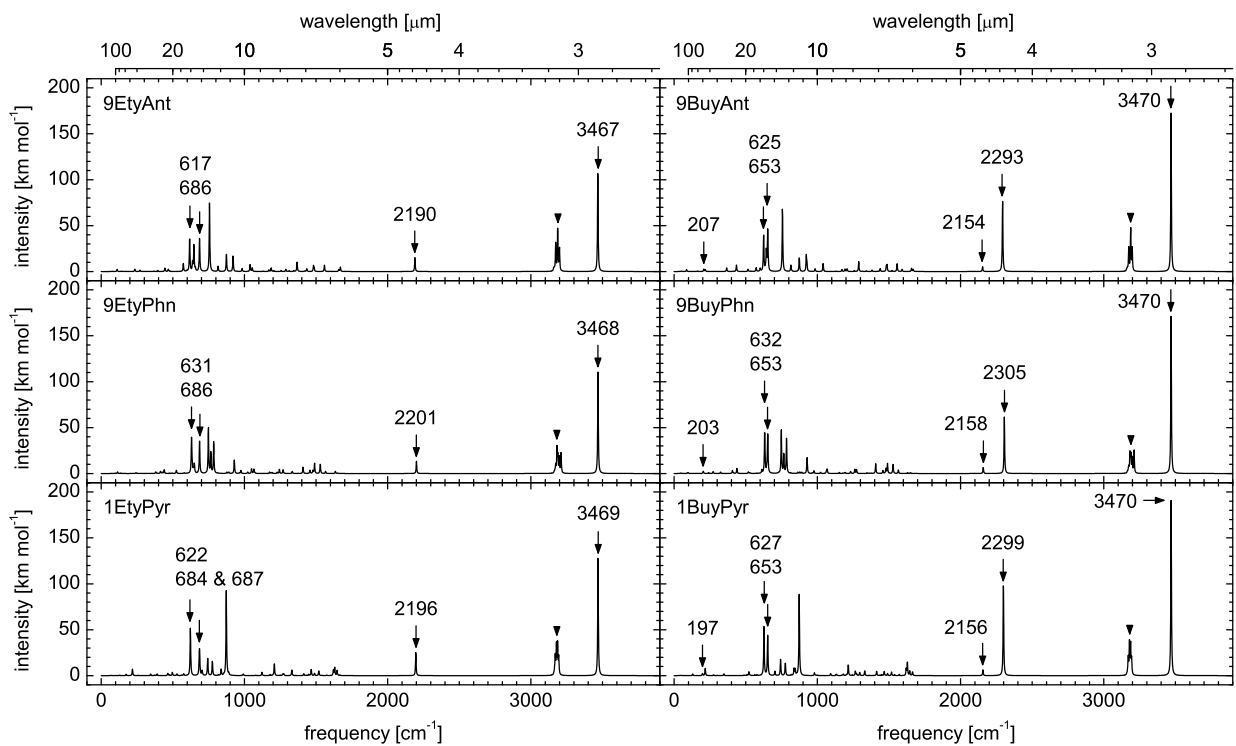}
\caption{Unscaled theoretical IR absorption spectra of the polyynyl-substituted PAHs computed using DFT methods. The bands have been given a Lorentzian profile with a FWHM of 5~cm$^{-1}$ as observed in the experimental spectra of Figure~\ref{fig5}. Arrows indicate bands due to characteristic vibrational modes of the side groups as described in Figure~\ref{fig5}. Aromatic CH stretching bands are labelled with full triangles.\label{fig6}}
\end{figure}

\begin{figure}
\epsscale{1.00}
% \plotone{fig7.eps}
\plotone{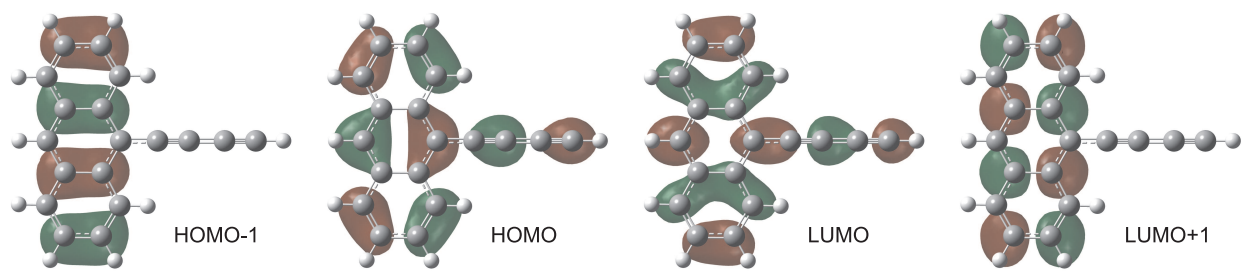}
\caption{Frontier orbitals of 9BuyAnt calculated at the TD-DFT-B3LYP/cc-pVTZ level of theory. For clarity, the foreground components of these $\pi$ orbitals are not shown. Similar results were obtained for 9EtyAnt and the substituted phenanthrenes. In the pyrenes, the four frontier orbitals involve both the PAH moiety and the side chain.\label{fig7}}
\end{figure}

\begin{figure}
\epsscale{0.50}
\plotone{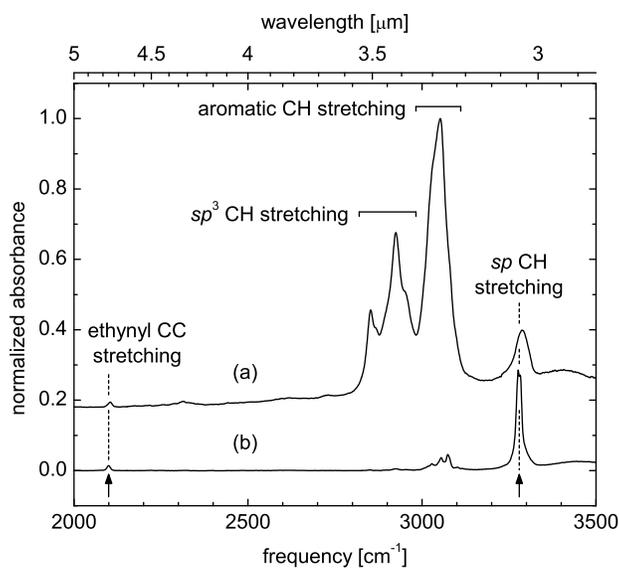}
\caption{Comparison of the IR spectrum of a soot extract in solution (sample CP85a) from \citet{Jaeger07} (a) with that of 9EtyPhn in a CsI pellet (b). Arrows and dashed lines indicate the bands attributed to the ethynyl CC stretching mode at $\sim$2095~cm$^{-1}$ (4.77~$\mu$m) and to the acetylenic CH stretching mode at $\sim$3280~cm$^{-1}$ (3.05~$\mu$m). Brackets indicate the bands due to the aliphatic (with $sp^3$-hybridized C) and aromatic CH stretching modes.\label{fig8}}
\end{figure}

\begin{figure}
\epsscale{0.50}
\plotone{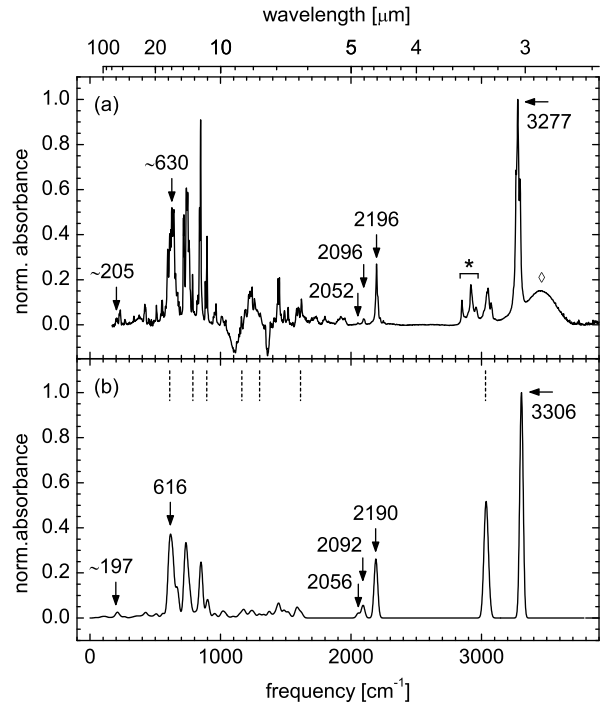}
\caption{Normalized sum of the experimental IR spectra shown in Figure~\ref{fig5} (a). Arrows labelled with frequencies indicate the collective features attributed to the C$\equiv$C$-$H out-of-plane and in-plane bending modes around 630~cm$^{-1}$ (15.9~$\mu$m), the butadiynyl antisymmetric CC stretching mode at 2052~cm$^{-1}$ (4.87~$\mu$m), the ethynyl CC stretching mode at 2096~cm$^{-1}$ (4.77~$\mu$m), the butadiynyl symmetric CC stretching mode at 2196~cm$^{-1}$ (4.55~$\mu$m), and the acetylenic CH stretching mode at 3277~cm$^{-1}$ (3.05~$\mu$m). The three peaks denoted by the asterisk correspond to CH stretching vibrations in groups where the C atom is $sp^3$-hybridized. The open diamond marks a broad feature due to water. Panel (b) displays the sum of the theoretical spectra of Figure~\ref{fig6} after frequency scaling. The bands were given a Gaussian profile with FWHM of 30~cm$^{-1}$. Vertical dashed lines mark the positions of the major AIBs~\citep{Tielens08}.\label{fig9}}
\end{figure}

Neither circumstellar nor interstellar polyynyl-substituted PAHs have been observed yet. The spectra of various IR sources do not show clear signs of their presence, neither as components of grains nor as free-flying molecules. According to the results of our IR measurements (Figure~\ref{fig5}) and theoretical calculations (Figure~\ref{fig6}), the frequencies of the acetylenic CCH bending and CH stretching modes do not vary greatly with the length of the polyynyl chain and the mass of the PAH moiety. As a consequence, collective contributions to the IR emission bands are expected from polyynyl-substituted PAHs. This is illustrated in Figure~\ref{fig9}, which represents, in one panel, the normalized sum of all the experimental IR spectra of Figure~\ref{fig5} and, in a second panel, the normalized sum of the theoretical IR spectra of Figure~\ref{fig6}. For a better comparison with the experimental sum spectrum, the theoretical vibrational frequencies have been scaled with two factors depending on their value, as described in Section~\ref{res:theo}. Moreover, in order to lend the theoretical absorption spectrum the aspect of an emission spectrum, the bands have been given a Gaussian profile with a FWHM of 30~cm$^{-1}$ \citep{Allamandola99b}.

First, the out-of-plane and in-plane acetylenic CCH bending modes would ideally give rise to two close features in the 16~$\mu$m region. It is however more likely that a single broad feature would be produced, as seen in Figure~\ref{fig9}, due to the slight variations of the wavelength positions from species to species and, considering emission spectra, due to the expected FWHM of $\sim$30~cm$^{-1}$ of each band \citep{Allamandola99b}. A search for this emission feature would be hindered by bands related to the deformation of aromatic rings which are present in the same wavelength region. As a fact, observations revealed a well-defined emission band at 16.4~$\mu$m \citep{Moutou00} and an underlying plateau between 15 and 20~$\mu$m \citep{VanKerckhoven00}, and both were assigned to CCC bending modes of PAH molecules \citep{Moutou00,VanKerckhoven00}. Still, a contribution of acetylenic CCH bending bands to the plateau cannot be ruled out.

Second, the acetylenic CH stretching modes of polyynyl-substituted PAHs would cause a band at $\sim$3280~cm$^{-1}$ (3.05~$\mu$m). To date, however, the only emission feature reported at a close wavelength is Pfund's hydrogen recombination band Pf$\epsilon$ at 3.04~$\mu$m.

Finally, as the number and energies of the acetylenic CC stretching modes vary depending on the polyynyl group, different PAHs carrying a specific group would produce collective features characteristic of that group. In ethynyl-subtituted PAHs, the unique acetylenic CC stretching mode gives rise to a weak band at $\sim$4.8~$\mu$m. The antisymmetric and symmetric acetylenic CC stretching modes of butadiynyl-substituted PAHs give two bands, a weak one at $\sim$4.9~$\mu$m and a medium one at $\sim$4.5~$\mu$m, respectively. There is no report in the literature of emission bands corresponding to acetylenic CC stretching modes. \citet{Duley09} suggested the possibility for CC stretching modes of $sp$-bonded chains, either acetylenic or cumulenic, to contribute to weak emission bands found at 5.25 and 5.7~$\mu$m \citep{Allamandola89a}, although a detailed study by \citet{Boersma09} attributed these emissions to various combinations of vibrations in pure non-substituted PAHs.

It should be mentioned that a search for IR emission bands of carbon chains, though not specifically as side groups, was undertaken by \citet{Allamandola99a}. Through the examination of the 5~$\mu$m region, where the bands of CC stretching modes are expected, they concluded that carbon chains in general contribute at best extremely little to the IR emission features.

While the features of carbon chains are not found in IR emission spectra, absorption bands observed toward IR sources have been attributed to acetylene \citep[C$_2$H$_2$, at 13.7~$\mu$m,][]{Lacy89}, diacetylene \citep[C$_4$H$_2$, at 8.0 and 15.9~$\mu$m,][]{Cernicharo01a,Cernicharo01b}, and triacetylene \citep[C$_6$H$_2$, at 8.1 and 16.1~$\mu$m,][]{Cernicharo01a,Cernicharo01b}. Another band of acetylene was observed at 2.44~$\mu$m \citep{Keady88}. An absorption feature measured at 3.0~$\mu$m, very close to the position of the acetylenic CH stretching band of polyynyl-substituted PAHs, was attributed to overlapping bands of hydrogen cyanide (HCN) and acetylene \citep{Chiar98,Goto03}.

Thus, IR observations do not disclose bands that can be assigned with certainty to polyynyl-substituted PAHs, neither in the regions that exhibit aromatic IR emission features, nor toward IR sources that reveal absorption bands of carbon chains. Still, the presence of polyynyl-substituted PAHs in circumstellar environments or in the ISM, i.e., in regions where conditions may allow their formation, even in low abundance, cannot be ruled out.

\subsubsection{Search for RF lines} \label{disc:astro:RF}

In general, non-substituted PAHs possess at best a very weak permanent electric dipole moment, supporting the theoretical values in the first line of Table~\ref{table1}. For this reason, laboratory measurements in the radiofrequency (RF) domain are not feasible, as reported by \citet{Thorwirth07} who attempted to measure the spectrum of Phn. Thus, a search for their RF lines is impracticable. On the contrary, our calculations indicate that polyynyl-substituted PAHs show a significant permanent electric dipole moment, which depends strongly on the length of the side chain according to the values in Table~\ref{table1}. Because these molecules possess a large moment of inertia, one must expect large partition functions and, as a consequence, weak lines that make detection difficult \citep{Tielens08}. As far as the polyynyl-substituted PAHs are concerned, laboratory measurements in the RF domain are missing, preventing a successful search for specific molecules with radiotelescopes. Such measurements can be envisaged, as demonstrated by the study of ethynylbenzene (phenylacetylene) by \citet{Cox75}.

\subsubsection{Search for UV/vis bands} \label{disc:astro:UVvis}

Assuming polyynyl-substituted PAHs are present in the diffuse ISM, there is a possibility to detect their absorption bands at UV/vis wavelengths. Bands of interstellar non-substituted PAHs were not discovered by searches in the UV/vis region, leading to conclude that such bands were either absent or too weak for detection at the given signal-to-noise ratio of the observations \citep{Clayton03,Gredel11,Salama11}. This conclusion is valid if one assumes that the bands carried by all PAHs present are separate. It is not valid, however, if one presumes that the bands overlap. In that case, they could give rise to a smooth contribution to the interstellar extinction curve, including its UV bump at 2175~{\AA} and its far-UV rise \citep{Steglich10,Steglich11a,Steglich12}. The strongest absorption bands of some of the polyynyl-substituted PAHs presented here (9BuyPhn, 1EtyPyr, and 1BuyPyr) are in the 3050--3850~{\AA} wavelength range within which a search for DIBs was fruitless \citep{Gredel11}. Thus, either their amount in the ISM is lower than the limit given by the signal-to-noise ratio of the observations or their features contribute to the interstellar extinction curve.

\subsubsection{Relation to the DIB carriers} \label{disc:astro:DIBs}

Even though the spectral features of the species we have studied do not coincide with observed emission or absorption bands, neither at IR nor at UV/vis wavelengths, they may point to a characteristic property of the elusive carriers of the DIBs, for which PAHs are candidate \citep{Crawford85,Leger85,vanderZwet85}. Non-substituted PAHs that show absorption bands at visible wavelengths, however, usually present features of significant intensity in the UV region as well. This is not consistent with the fact that DIB carriers apparently exhibit a single strong band which is found at a visible wavelength. Moreover, the PAH family, even limited to the non-substituted PAHs, comprises, so to say, an infinite number of species. In contrast, as noted by \citet{Salama96}, the number of DIBs of moderate to strong intensity is about 50. Because these DIBs do not show any evidence of correlated variations when observed toward various lines of sight, the existence of a specific carrier for each DIB has been inferred. Assuming that the carriers are PAH molecules, they must stand out among the members of their family. In that case, criteria such as size distribution and charge state may not be sufficient to limit the number of potential carriers. Another selection mechanism, based on the formation and destruction processes of the carriers, was discussed by \citet{Duley06}.

Led by the present study, we may propose that DIB carriers are characterized by specific combinations of aromatic cores with side groups. These PAH molecules carrying side groups would be intermediate products in the destruction or formation routes of interstellar PAHs. The number of DIB carriers would be limited due to the selection mechanism imposed by their formation and destruction processes and by the specificity of the combinations of aromatic cores with side groups. Due to the side groups, electronic transitions at UV wavelengths would give rise to broadened features that would blend in the interstellar extinction curve, while transitions in the visible region would be associated with well-defined bands. Their oscillator strength would be greater than those of the parent PAHs since more electrons are involved. It could also be that another ingredient has to be considered in order to obtain a molecule that can carry a strong DIB since side groups may not be sufficient to obtain transitions in the vis to near IR spectral regions.

The DIBs correlate neither with the AIBs, attributed to PAH molecules, nor with the UV bump at 2175~{\AA} \citep{Xiang11}, to which interstellar PAHs may contribute (see \citealt{Steglich10,Steglich11b} and references therein). This could be construed as an argument against PAHs being carriers of the DIBs. As an answer to such an interpretation, we may consider that two populations of PAH molecules exist in space. The first population would be produced in carbon-rich circumstellar envelopes through the HACA mechanism. This population would essentially consist of a great diversity of non-substituted PAHs. It would be denser close to the stars, giving the AIBs, and very diffuse in the ISM, causing a smooth contribution to the UV extinction curve, the population of each species being too low for individual detection. The second population would be produced in molecular clouds by the EAM phenomenon and would consist of PAHs with side groups, some of which being DIB carriers.

\section{Conclusion}

We have examined the relevance of polyynyl-substituted PAH molecules to astrophysics. It appears that ethynyl-substituted PAHs may form in circumstellar envelopes through the HACA mechanism. Polysubstituted PAHs carrying ethynyl and butadiynyl groups may form in molecular clouds, even cold ones, through neutral-neutral reactions of PAHs with ethynyl radicals (EAM). As polyynyl radicals have been identified in space, it would be interesting to know whether they can react with PAHs as easily as ethynyl, which would result in the formation of polyynyl-substituted PAHs. One can expect that the photostability of polyynyl-substituted PAHs is lower than that of the non-substituted PAHs. Due to conjugation of the $\pi$ orbitals over both the PAH and polyynyl moieties, the molecules most likely withstand the UV radiation field. Moreover, the dramatic broadening that we have observed for the absorption bands in the UV region suggests a very short lifetime of the electronically excited states, which is attributed to fast internal conversion. Therefore, it is reasonable to assume the existence of stationary populations of polyynyl-substituted PAHs in regions where PAH growth takes place.

Like the populations of non-substituted PAHs, populations of polyynyl-substituted PAHs would cause collective features in the IR wavelength region, as illustrated by the spectra we have measured and calculated. Since the observation of these features has not been reported yet, it can be concluded that the populations are below the detection level at IR wavelengths. It is the case in IR sources, where IR emission bands of non-substituted PAHs are observed, and nearby these sources, toward which IR absorption bands of hydrocarbons are observed.

In the UV/vis region, the molecules would not give rise to collective features. Nevertheless, as the absorption cross section is in general larger for electronic transitions than for vibrational transitions, searches for individual species can be proposed. Although the present spectra were measured in rare gas matrices and hence cannot be directly compared with interstellar spectra, all the bands we have observed would lie at UV wavelengths under the conditions of the ISM. Since surveys of the UV region have not revealed any band that could be attributed to a PAH molecule, we conclude that the amount of each of the species we have studied is too small for detection with the current limitations.

Our most interesting observation concerns the intensity profile of the absorption bands in the electronic spectra. The bands in the upper UV range are dramatically broadened while those in the lower range, arising from the first two electronic transitions, are barely affected. We interprete this broadening of the bands with shorter wavelengths as the effect of very fast deexcitation through internal conversion, in which the polyynyl group plays a crucial role. Such broad bands would be difficult to identify in interstellar spectra and would only contribute to the extinction curve of the ISM. On the other hand, the strongest band in the spectra of the first two electronic transitions may stand out. Thus, the electronic spectra of polyynyl-substituted PAHs present similarities with the spectra of the DIB carriers. Although the molecules we have studied are not DIB carriers, our study may represent a new step toward their identification by introducing the cause for the lack of DIBs in the UV region and the limited number of strong DIBs.

\acknowledgments

This work was carried out within a cooperation between the Max-Planck-Institut f\"ur Astronomie, the Friedrich-Schiller-Universit\"at Jena, and the Technische Universit\"at Dresden. The support of the Deutsche Forschungsgemeinschaft (contract No. Hu 474/24) is gratefully acknowledged. We are thankful to Dr. H. Mutschke for giving us access to the FTIR spectrometer and to G. Born for preparing the CsI pellets and performing the HPLC measurements. Our thanks also go to the anonymous referee for suggestions that helped to improve the quality of this article.

\clearpage

\clearpage

\clearpage

\clearpage

\clearpage

\clearpage

\clearpage

\clearpage

\clearpage

\clearpage

\begin{table}
\begin{center}
\caption{Theoretical Permanent Electric Dipole Moments\label{table1}}
\begin{tabular}{llll}
\tableline
\tableline
$R$ & 9$R$Ant & 9$R$Phn & 1$R$Pyr \\
\tableline
H$-$ & 0.0000 & 0.0257 & 0.0000 \\
HCC$-$ & 0.5893 & 0.7466 & 0.9251 \\
HCCCC$-$ & 1.3158 & 1.4478 & 1.7736 \\
\tableline
\end{tabular}
\tablecomments{These values, expressed in units of Debye, were obtained for the electronic ground state geometry determined at the DFT-B3LYP/cc-pVTZ level of theory.}
\end{center}
\end{table}

\clearpage

\begin{table}
\begin{center}
\caption{Electronic States and Transitions\label{table2}}
\begin{tabular}{llllllll}
\tableline
\tableline
Molecule & Transition\tablenotemark{a} & \multicolumn{2}{c}{ZINDO} & \multicolumn{2}{c}{TD-DFT} & \multicolumn{2}{c}{Observed} \\
 & & $\lambda$/{\AA} & Strength & $\lambda$/{\AA} & Strength & $\lambda$/{\AA} & Assignment \\
\tableline
Ant     & 1 & 3691 & 0.2222 & 3855 & 0.0525 & 3634 & $S_1 \leftarrow S_0$ \\
        & 2 & 2631 & 2.4307 & 2372 & 1.9722 & 2374 & $\beta$-band \\
9EtyAnt & 1 & 3984 & 0.3314 & 4133 & 0.1096 & 3868 & $S_1 \leftarrow S_0$ \\
        & 2 & 2692 & 2.1803 & 2466 & 1.5623 & 2440 & $\beta$-band \\
9BuyAnt & 1 & 4212 & 0.4358 & 4329 & 0.2157 & 3998 & $S_1 \leftarrow S_0$ \\
        & 2 & 2718 & 2.0456 & 2547 & 1.1209 & 2490 & $\beta$-band \\
Phn     & 1 & 3096 & 0.1424 & 2937 & 0.0580 & 2844 & $S_2 \leftarrow S_0$ \\
        & 2 & 3335 & 0.0011 & 3113 & 0.0015 & 3412 & $S_1 \leftarrow S_0$ \\
9EtyPhn & 1 & 3282 & 0.3114 & 3139 & 0.1878 & 3010 & $S_2 \leftarrow S_0$ \\
        & 2 & 3400 & 0.0029 & 3238 & 0.0004 & 3504 & $S_1 \leftarrow S_0$ \\
9BuyPhn & 1 & 3621 & 0.2794 & 3371 & 0.3635 & 3156 & $S_2 \leftarrow S_0$ \\
        & 2 & 3427 & 0.0028 & 3327 & 0.0470 & 3544 & $S_1 \leftarrow S_0$ \\
Pyr     & 1 & 3526 & 0.5175 & 3366 & 0.2518 & 3236 & $S_2 \leftarrow S_0$ \\
        & 2 & 3629 & 0.0027 & 3300 & 0.0003 & 3674 & $S_1 \leftarrow S_0$ \\
1EtyPyr & 1 & 3747 & 0.6753 & 3621 & 0.4012 & 3438 & $S_2 \leftarrow S_0$ \\
        & 2 & 3707 & 0.0107 & 3417 & 0.0012 & 3776 & $S_1 \leftarrow S_0$ \\
1BuyPyr & 1 & 3960 & 0.7316 & 3851 & 0.6080 & 3578 & $S_2 \leftarrow S_0$ \\
        & 2 & 3736 & 0.0063 & 3495 & 0.0047 & 3820 & $S_1 \leftarrow S_0$ \\
\tableline
\end{tabular}
\tablenotetext{a}{Transition type 1: (HOMO $\rightarrow$ LUMO and HOMO$-1 \rightarrow$ LUMO$+1$). Transition type 2: (HOMO $\rightarrow$ LUMO$+1$ and HOMO$-1 \rightarrow$ LUMO).}
\tablecomments{The theoretical values were obtained with the ZINDO and TD-DFT-B3LYP/cc-pVTZ models applied at the electronic ground state geometry determined at the DFT-B3LYP/cc-pVTZ level of theory.}
\end{center}
\end{table}

\end{document}